\documentclass[preprint,3p,authoryear,twocolumn]{elsarticle}
\usepackage{graphicx}
\usepackage{amssymb}
\usepackage{lineno}
\usepackage{tabularx}
\journal{Journal of Atmospheric and Solar-Terrestrial Physics}
\begin{document}
\begin{frontmatter}
\title{SPECIFIC EVANESCENT ACOUSTIC-GRAVITY MODES IN ISOTHERMAL ATMOSPHERE}
\author{A. K. Fedorenko}
\ead{fedorenkoak@gmail.com}
\author{Yu. O. Klymenko$^*$}
\ead{yurklym@gmail.com}
\cortext[cor1]{Corresponding author.}
\author{O. K. Cheremnykh}
\ead{oleg.cheremnykh@gmail.com}
\author{E. I. Kryuchkov}
\ead{kryuchkov.ye@gmail.com}

\address{Space Research Institute, prosp. Akad. Glushkova 40, build. 4/1, 03187, Kyiv, Ukraine}
\begin{abstract}
We demonstrate the existence a family of four specific evanescent wave modes in an isothermal atmosphere. These modes are the solutions of the linearized system of hydrodynamic equations with respect to perturbed quantities -- horizontal and vertical components of velocity, density, and temperature provided that one of them equals zero. Such condition excludes the free propagation and leads to the specific evanescent modes, i.e. the  Lamb wave, the non-propagating Brunt-V\"{a}is\"{a}l\"{a} oscillations, the non-divergent (f-mode), and the inelastic ($\gamma $-mode). On this basis, such indicated modes form the family of evanescent modes in the isothermal atmosphere. The polarization relations for the evanescent modes are also included and analyzed. The possibility of identifying these modes in observations on the Sun and in the Earth's atmosphere is considered.
\end{abstract}
\end{frontmatter}
%\linenumbers

\section{INTRODUCTION}

The theory of acoustic gravity waves (AGWs) admits the existence of continuous spectrum of freely propagating waves, as well as horizontally propagating (evanescent) wave modes in an isothermal atmosphere. In the physics of the Sun and planetary atmospheres, the main focus is on the study of freely propagating AGWs (Hines, 1960; Tolstoy, 1963; Stenflo and Shukla, 2009; Vadas and Fritts, 2005; Zhang and Yi, 2002). At the same time, the role of evanescent wave modes among other atmospheric wave processes has been studied to a much lesser extent (Jones, 1969; Waltercheid and Hecht, 2003; Roy et al., 2019; Cheremnykh et al., 2019).

Unlike the freely propagating waves, the vertical component of the wave vector for evanescent solutions is a purely imaginary quantity which determines the exponential changes of the amplitudes in the vertical direction. In an infinite atmosphere, the energy of the evanescent wave has to be decreased on both sides of the propagation level. It requires the suitable surfaces in the medium (discontinuity of atmospheric parameters) on which such condition can be fulfilled. Moreover, the polarization relations between fluctuations of different perturbed quantities (velocity, pressure, density, and temperature) may contain singularities for evanescent solutions. The possibility of realizing the evanescent waves with a mathematical singularity was considered in (Cheremnykh et al., 2021).

In this paper, we show that in the isothermal atmosphere there is a family of four specific  evanescent modes. Three of them have been well-known earlier. These are the Brunt-V\"{a}is\"{a}l\"{a} oscillations, the Lamb wave, and the surface f-mode (Lamb, 1932; Jones, 1969; Gossard and Hooke, 1975; Waltercheid and Hecht, 2003). However, the recent discovery of the $\gamma $-mode (Cheremnykh et al., 2019) made it possible to consider all these modes from a common position. We will analyze the possibility of identifying these evanescent modes in observations on the Sun and in the Earth's atmosphere using polarization relations. Special attention will be paid to the $\gamma $-mode, which has not yet been experimentally observed.

\section{BASIC EQUATIONS}

In the model of an infinite isothermal atmosphere, the small perturbations of the medium are usually described by well-known linearized system of hydrodynamic equations. For perturbations of the horizontal ($V_{x} $) and vertical components ($V_{z} $) of the particle velocity, relative density ($\rho'/\rho _{0} $) and temperature ($T'/T_{0} $) fluctuations of the atmosphere, this system can be written as (Kundu, 1990):
\begin{equation} \label{1}
\begin{array}{l}
\frac{\partial V_{x} }{\partial t} +gH\frac{\partial }{\partial x} \left(\frac{\rho '}{\rho _{0} } +\frac{T'}{T_{0} } \right)=0,\\
\frac{\partial V_{z} }{\partial t} +gH\frac{\partial }{\partial z} \left(\frac{\rho '}{\rho _{0} } +\frac{T'}{T_{0} } \right)-g\frac{T'}{T_{0} } =0,\\
\frac{\partial }{\partial t} \left(\frac{T'}{T_{0} } \right)+\left(\gamma -1\right){\rm div}\vec{V}=0,\\
\frac{\partial }{\partial t} \left(\frac{\rho'}{\rho_{0} } \right)+\frac{1}{\rho_{0} } \frac{\partial \rho _{0} }{\partial z} V_{z} +{\rm div}\vec{V}=0.
\end{array}
\end{equation}
Here $\rho_{0} $ and $T_{0} $ are the density and the temperature of the medium, $H=kT_{0}/mg$ is the atmosphere scale height, $k$ is the Boltzmann constant, $m$ is the mass of a particle of atmospheric gas, $g$ is the acceleration of gravity, and $\gamma $ is specific heat ratio.

The equation system (\ref{1}) has been obtained for an isothermal atmosphere which is barometrically stratified in the gravity field,
\[\frac{1}{\rho _{0} } \frac{\partial \rho _{0} }{\partial z} =-\frac{1}{H},\]
and includes the equations of motion, the heat balance and the continuity. The equations are written in the two-dimensional Cartesian coordinate system with vertical axis $z$ directed upward and horizontal axis $x$ along the particle velocity vector.

Equation system (\ref{1}) describes different types of linear perturbations that can exist in an isothermal atmosphere. For high frequencies and small wavelengths, the spectrum of these perturbations is limited by the collision frequency and the mean free path of particles. Note that in an atmosphere with exponentially decreasing density, these restrictions are quickly increasing with the altitude. For low frequencies and long wavelengths, the spectrum is limited by periods of several hours and horizontal scales of $\sim$ 1--2 thousand km. The time limitations are associated with the effect of the Earth's rotation and the spatial constraints are connected to the Earth's  curvature.

\section{FREELY PROPAGATING AND EVANESCENT WAVES}

System (\ref{1}) implies two types of wave disturbances differing in the way they propagate in the atmosphere. The first type is acoustic and gravitational waves which freely propagate at an angle to the horizontal plane. The second type is horizontally propagating (evanescent) waves.

We will seek the solution of system (\ref{1}) in general form, without specifying the dependence on the vertical coordinate:
%\begin{equation}\label{2}
%\!\!\!\!\!\!\!\!\!T'/T_{0} ,\rho'/\rho _{0},V_{x} ,V_{z} \sim \exp(az)\exp\left[i\left(\omega t-k_{x} x\right)\right]=0,
\begin{equation}\label{2}
T'/T_{0},\rho'/\rho _{0},V_{x},V_{z}\sim e^{az}e^{i\left(\omega t-k_{x}x\right)},
\end{equation}
where $\omega $ is the frequency, $k_{x} $ is the horizontal component of the wave vector, and constant value of $a$ determines the height dependence of the amplitude. In general case, it can take a complex value.

After substituting solution (\ref{2}) into system (\ref{1}), we obtain the following equation for the quantity of $a$:
\begin{equation}\label{3}
a^{2} -\frac{a}{H} +k_{x}^{2} \left(\frac{N^{2} }{\omega ^{2} } -1\right)+\frac{\omega ^{2} }{c_{s}^{2} } =0
\end{equation}
with solutions
\begin{equation}\label{4}
\!\!\!\!\!a=\frac{1}{2H} \pm \left[\frac{1}{4H^{2} } +k_{x}^{2} \left(1-\frac{N^{2} }{\omega ^{2} } \right)-\frac{\omega ^{2} }{c_{s}^{2} } \right]^{1/2}.
\end{equation}
Here $N^{2} =g\left(\gamma -1\right)/\gamma H$ is the square of the Brunt-V\"{a}is\"{a}l\"{a} (BV) frequency.

Taking into account expressions (\ref{2}) and (\ref{4}), we can see that the wave disturbances are divided into freely propagating and evanescent ones depending on the sign of expression under the radical in Eq. (\ref{4}). If the expression is negative, then $a$ is a complex quantity $a=\left(1/2H\right)\pm ik_{z} $, where $k_{z} $ the is the vertical component of the wave vector. For such waves, the real part of the quantity $a$ determines the amplitude change in the vertical direction, providing the wave energy conservation with the height. The considered case corresponds to the freely propagating acoustic waves with frequency $\omega ^{2} >c_{s}^{2} /4H^{2} $ and the gravitational waves with $\omega ^{2} <N^{2} $ (Hines, 1960). Different signs of the imaginary part of $a$ indicate the direction of wave propagation (up or down).

If the expression under the radical is positive, then the waves can propagate only in the horizontal plane. In this case, $a=\left(1/2H\right)\pm \kappa $, where $\kappa $ is a real value, describes a change in their amplitudes with the height. Different signs of the value of $\kappa $ mean that a pair of evanescent solutions always corresponds to the same dispersion. They are called as the wave and the pseudo-wave (or "mode" and "pseudo-mode"). Usually, a wave mode is understood as a solution that is obtained under certain assumptions regarding the physical properties of perturbations. And a pseudo-mode is a solution that mathematically complements the direct mode. Fully coinciding in the dispersion, they differ in the polarization and in the amplitude dependence on the height.

It is not easy to distinguish a mode from a pseudo-mode in experimental data, since they have the same dispersion. For example, the mode with dispersion $\omega ^{2} =k_{x} g$ observed on the Sun is usually identified with the f-mode. It is diagnosed by simultaneous measurements of the frequency and the horizontal wavelength. To distinguish the mode from the pseudo-mode, it is needed  some additional information on polarization, which is not always possible to do using modern observation methods.

\section{SPECIFIC EVANESCENT MODES}

Evanescent modes usually follow from the system of linearized hydrodynamic equations at some conditions which are imposed on the perturbations (Cheremnykh et al, 2019). Let us show that system (\ref{1}) contains four special solutions, which are obtained by vanishing one of the four perturbed quantities ($T'{\rm /}T_{0} $,$\rho '/\rho _{0} $,$V_{x} $, or $V_{z} $). Considering $V_{x} =0$, we obtain the well-known solution in the form of non-propagating Brunt-V\"{a}is\"{a}l\"{a} oscillations: $\omega ^{2} =N^{2} $, $a=1/\gamma H$ (Gossard and Hooke, 1975). If $V_{z} =0$, then we get the Lamb wave $\omega ^{2} =k_{x}^{2} c_{s}^{2} $, $a=\left(\gamma -1\right)/\gamma H$ (Lamb, 1932). For $T'{\rm /}T_{0} =0$, it implies the f-mode: $\omega ^{2} =k_{x} g$, $a=k_{x} $. This solution is also called the non-divergent mode, since it was first derived from the condition $div\vec{V}=0$ (Jones, 1969). For $\rho '/\rho _{0} =0$, the $\gamma $-mode follows from system (\ref{1}): $\omega ^{2} =k_{x} g\left(\gamma -1\right)$, $a=\left(1/H\right)-k_{x} $. It was first obtained from ${\rm div}(\rho _{0}\vec{V})=0$, that's  why is also called the inelastic mode (Cheremnykh et al, 2019).

The location of these four modes is shown on the spectral diagram (see Fig. 1). Their basic properties are summarized in Table 1. As can be seen from the Table, the polarization relations for all these modes have a very simple form. Some relations between the perturbed quantities are independent on the wave spectral properties. So, when one of the perturbed thermodynamic quantities ($T'{\rm /}T_{0} $ or $\rho '{\rm /}\rho _{0} $) is equal to zero, then the relation between the velocity components becomes independent on the spectral properties. If one of the velocity components is equal to zero, then the polarization relation between the temperature and the density fluctuations is also independent on $\omega $ and $k_{x} $. For example, regardless of the wave spectral properties, non-divergent and inelastic modes have the clockwise and counter-clockwise circular polarizations. In the BV oscillations, the motions occur only vertically. In the Lamb wave, it occurs in the horizontal plane.

\begin{table*}[t]
  \centering
\caption{Characteristics of specific evanescent modes in isothermal atmosphere}
%\begin{tabular}{|c|c|c|c|c|} \hline
\begin{tabular}{|p{0.9in}|p{0.9in}|p{1.in}|p{1.5in}|p{1.1in}|} \hline
Type $\newline$ of mode &Non-divergent or f-mode  & Inelastic or \newline $\gamma$-mode  &  BV oscillations &  Lamb wave \\ \hline
Condition for\newline obtaining & $T'{\rm /}T_{0} =0$,\newline ${\rm div}\vec{V}=0$ & $\rho'/\rho_{0} =0$,\newline ${\rm div}(\rho _{0} \vec{V})=0$ & $V_{x} =0$ & $V_{z} =0$ \\ \hline
Dispersion  & $\omega ^{2} =k_{x} g$ & $\omega ^{2} =k_{x} g\left(\gamma -1\right)$ & $\omega^{2} =g(\gamma -1)/\gamma H=N^{2} $ & $\omega ^{2} =k_{x}^{2} c_{s}^{2} $ \\ \hline
Vertical dependence of amplitude  & $\newline$ $a=k_{x} $ & $\newline$  $a=\left(1/H\right)-k_{x} $ & $\newline$  $a=1/\gamma H$ & $\newline$  $a=(\gamma -1)/\gamma H$ \\ \hline
$\newline$  Polarization & $V_{x} =\frac{k_{x} }{\omega } gH\frac{\rho '}{\rho _{0} }, $\newline $V_{z} =i\omega H\frac{\rho '}{\rho _{0} }, $\newline $V_{z} =iV_{x} $ & $V_{x} =\frac{k_{x} }{\omega } gH\frac{T'}{T_{0} }, $\newline $V_{z} =-i\,\frac{\omega H}{\gamma-1}\, \frac{T'}{T_{0} } $,\newline $V_{x} =iV_{z} $ & $\frac{\rho '}{\rho _{0} } =i\,\frac{\gamma -1}{\gamma H}\, \frac{V_{z} }{\omega } $,\newline $V_{z} =-i\,\frac{g}{\omega }\, \frac{T'}{T_{0} } $,\newline $\frac{\rho '}{\rho _{0} } =-\frac{T'}{T_{0} } $; $\frac{p'}{p_{0} } =0$ & $\frac{\rho '}{\rho _{0} } =\frac{k_{x} }{\omega }\, V_{x} $,\newline $\frac{T'}{T_{0} } =\frac{k_{x} }{\omega }(\gamma -1)V_{x}, $\newline $\frac{T'}{T_{0} } =\left(\gamma -1\right)\frac{\rho '}{\rho _{0} } $ \\ \hline
\end{tabular}
\end{table*}

With the exception of the inelastic mode $\omega ^{2} =k_{x} g(\gamma -1)$, which was recently discovered (Cheremnykh et al, 2019), the rest of the evanescent solutions have been studied separately for several decades. To obtain them, the authors imposed various restrictions on the components of the particle velocity vector in system of equations (\ref{1}): 1) the Lamb wave was obtained under  assumption $V_{z} =0$; 2) BV oscillations occur at $V_{x} =0$; 3) non-divergent mode provided at $div\vec{V}=0$;  4) inelastic mode was obtained at $div\left(\rho _{0} \vec{V}\right)=0$. At the same time, the fluctuations of thermodynamic quantities ($T'{\rm /}T_{0} $,$\rho'/\rho _{0} $) for these evanescent modes were analyzed weakly.

%%%%%%%%%%%%%%%%%%%
\begin{figure}[b!]
\centering
\includegraphics[width=\columnwidth]{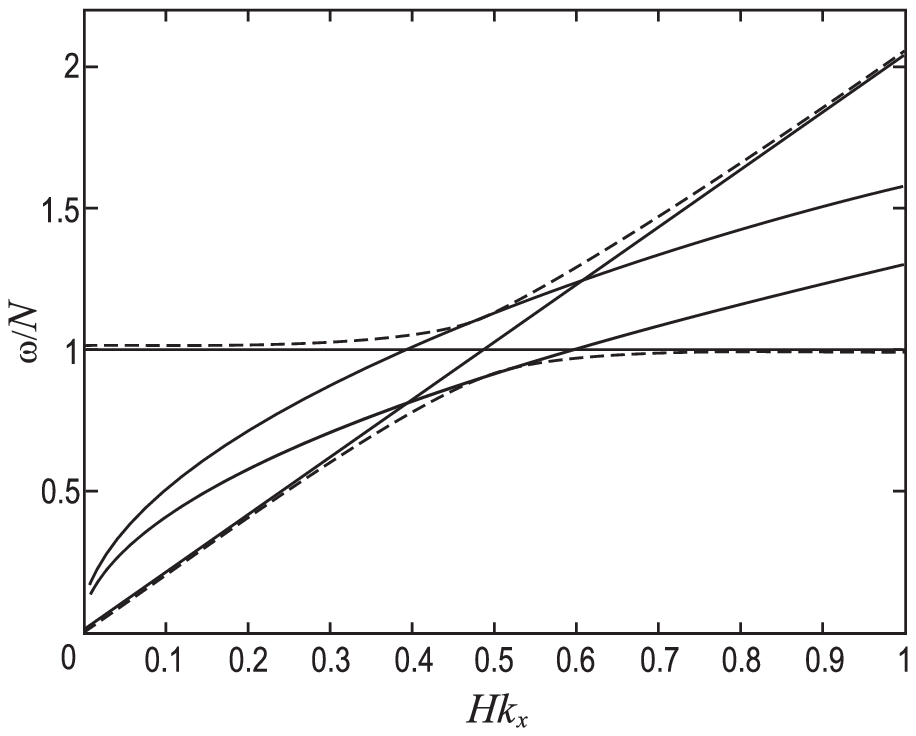}
\caption{Dispersion dependencies $\omega \left(k_{x} \right)$  for four specific evanescent wave modes in the isothermal atmosphere. Dashed lines determine the boundaries between evanescent and freely propagating wave regions. Dispersion dependences $\omega =\sqrt{k_{x} g}$  and $\omega =\sqrt{k_{x} g\left(\gamma -1\right)} $  correspond to the upper and the lower solid curves, respectively. Dispersions $\omega =N$ and $\omega =k_{x} c$ are shown by the horizontal and  the sloping straight lines, respectively.}
%\label{fig:1}
\end{figure}
%%%%%%%%%%%%%%%%%%%

We notice for the first time that all these modes can be obtained in the same way from the system of four first-order equations (\ref{1}) by equating to zero one of the four perturbed quantities $T'{\rm /}T_{0} $, $\rho '/\rho _{0} $, $V_{x} $, or $V_{z} $. In this sense, these solutions form a certain family of evanescent modes of the isothermal atmosphere, which remained incomplete until the discovery of the $\gamma $-mode with $\rho '/\rho _{0} =0$. Note additionally, that vanishing of any of the perturbed quantities in system (\ref{1}) is the condition that excludes the possibility of the free wave propagation and leads only to the evanescent solutions.

Note, that the considered modes do not interact at points of the intersection due to the difference in polarizations (see Table 1). There are five intersection points of the specific evanescent modes and two points where the $\gamma $ and f- modes touch the regions of freely propagating AGWs (see Fig. 1). The impossibility of interaction of evanescent modes at the intersection points is also clear from the following mathematical reasoning. Since one of the quantities ($T'{\rm /}T_{0} $,$\rho '/\rho _{0} $,$V_{x} $ or $V_{z}$) vanishes on each of the four specific dispersion curves, two perturbed quantities must be equal to zero at the point of intersection of these two curves. But, each of the equations of system (\ref{1}) contains only three perturbed quantities, therefore, the requirement that two quantities equal zero means that the third quantity vanishes too.

\section{POLARIZATION RELATIONS FOR EVANESCENT WAVEW MODES}

The general features of the polarization of evanescent waves are clear from simple energy considerations. Since such waves propagate only in the horizontal plane, no energy transfer in the vertical direction should occur. Consequently, the vertical wave energy flux (averaging over the period) $\bar{F}_{z} =\left\langle p'\cdot \vec{V}\right\rangle _{z} =0$ (where $p'$ is the perturbed pressure), as well as the average vertical momentum $\bar{P}_{z} =\left\langle \rho '\cdot \vec{V}\right\rangle _{z} =0$. The requirement the vanishing the vertical energy imposes general restrictions on the polarization relations for evanescent waves. For an ideal atmospheric gas where $p'/p_{0} =\left(T'/T_{0} \right)+\left(\rho '/\rho _{0} \right)$, it follows from the first equation (\ref{1}) that $p'=\rho_{0} \left(\omega /k_{x} \right)V_{x} $. Therefore, the condition $\bar{F}_{z} =0$ is actually equivalent to the fact that the oscillations of the vertical and horizontal velocity components are phase-shifted by $\pm \pi /2$ in evanescent waves. Deviation from this shift means that there is a transfer of the wave energy in the vertical direction.

For freely propagating AGWs, the phase shift between the oscillations of the vertical and the horizontal velocity components are varied in the interval $\left(0;\pm \pi /2\right)$ depending on the spectral properties (Hines, 1960). Such waves, propagating at an angle to the horizontal plane, transfer the energy in the vertical direction. It should be noted that the vertical flux of AGW energy has a pronounced maximum at certain spectral parameters (Kryuchkov and Fedorenko, 2012). Consequently, such waves are most efficient for an energy transport between different altitudes of the atmosphere.

Since the oscillations of $V_{x} $ and $V_{z} $ for evanescent waves are phase-shifted by $\pm \pi /2$,  the value of ${\rm div}\vec{V}=\left(\partial V_{x} /\partial x\right)+\left(\partial V_{z} /\partial z\right)\sim V_{z} $ at real $a$. Then, from the third and the fourth equations of system (\ref{1}) it follows that $T'{\rm /}T_{0} $ and $\rho '/\rho _{0}$ are also phase-shifted by  $\pm \pi /2$ relative to $V_{z} $. This is also in agreement with condition $\bar{P}_{z} =0$.

According to satellite observations, AGWs dominate in the polar thermosphere, in which oscillations of $\rho '/\rho _{0}$ and $V_{z} $ occur almost in the phase (Fedorenko and Zakharov, 2012). This feature indicates the effective momentum transfer in the vertical direction and can be used for excluding the evanescence of observed waves.

Let us obtain the polarization relations for AGWs from equation system (\ref{1}). For clarity, we express all perturbed quantities in terms of variable $V_{z} $:
\begin{equation}\label{5}
V_{x} =iV_{z} \left[\frac{k_{x} c_{s}^{2} \left(a-1/\gamma H\right)}{\omega ^{2} -k_{x}^{2} c_{s}^{2} } \right],
\end{equation}
\begin{equation}\label{6}
\frac{p'}{p_{0} } =iV_{z} \left[\frac{\gamma \omega \left(a-1/\gamma H\right)}{\left(\omega ^{2} -k_{x}^{2} c_{s}^{2} \right)} \right],
\end{equation}
\begin{equation}\label{7}
\frac{T'}{T_{0} } =iV_{z} \left[\frac{\left(\gamma -1\right)\left(a\omega ^{2} -k_{x}^{2} g\right)}{\omega \left(\omega ^{2} -k_{x}^{2} c_{s}^{2} \right)} \right],
\end{equation}
\begin{equation}\label{8}
\frac{\rho '}{\rho _{0} } =iV_{z} \left[\frac{k_{x}^{2} g\left(\gamma -1\right)+\omega ^{2} \left(a-1/H\right)}{\omega \left(\omega ^{2} -k_{x}^{2} c_{s}^{2} \right)} \right].
\end{equation}

Written in general form polarization relations (\ref{5})-(\ref{8}) are valid for both evanescent and freely propagating waves. As can be seen from these expressions, the phase shifts between the perturbed quantities for evanescent waves (here $a$ is a real) do not depend on the spectral characteristics. Only the amplitude ratios of evanescent disturbances depend on  $\omega $, $k_{x} $ and $a$, as well as on the parameters of the medium. For freely propagating AGWs we have $a=\left(1/2H\right)\pm ik_{z} $, and therefore, the phase shifts between the perturbed quantities, as well as the ratios of their amplitudes, depend on the spectral characteristics. Also, none of the perturbed quantities can be equal to zero in freely propagating AGWs.

Polarization relations (\ref{5})-(\ref{8}) are valid in the entire spectral region of AGWs. The peculiarity in the denominator of expressions (\ref{5})-(\ref{8}) arises from the condition $V_{z} =0$ on dispersion of the Lamb wave $\omega ^{2} =k_{x}^{2} c_{s}^{2} $. When obtaining the polarization relations for the four specific evanescent modes, it is convenient to equate zero the corresponding perturbed quantity in system (\ref{1}). After that, polarization relations are obtained without any peculiarities (see Table 1).

The perturbed quantity, which vanishes on the dispersion curve, undergoes a phase discontinuity by $\pi $ when passing through it. The rest of the perturbed quantities have no peculiarities near this curve. It means that the phase shifts between the pairs of the corresponding quantities change abruptly by $\pi $ on opposite sides of the dispersion curve. For example, for the Lamb curve, these will be the phase shifts between pairs of perturbed quantities $V_{x} -V_{z} $, $\left(p'/p_{0} \right)-V_{z} $, $\left(T'/T_{0} \right)-V_{z} $, $\left(\rho '/\rho _{0} \right)-V_{z} $ which will change abruptly from $+\pi /2$ to $-\pi /2$, or vice versa. It is important to understand these features of polarization when interpreting observations of the evanescent spectrum of AGWs, for example, on the Sun.

\section{EVANESCENT MODES IN OBSERVATIONS}

When analyzing the wave processes in observational data, it is important to determine the type of the wave. To distinguish freely propagating waves from evanescent ones, it is sufficient to analyze the polarization relations between the perturbed quantities, if experimental data allow doing it. The jumps in the phase between individual pairs of the perturbed quantities are the characteristic of evanescent waves only. At the same time, the phase shifts are changed monotonically in the polarization relations for freely propagating AGWs.

Polarization relations indicate discontinuities in phase shifts between pairs of perturbed quantities on those dispersion curves where one of these quantities is equal to zero. Therefore, in the experimental study of different types of evanescent modes, the choice of the observed variables is important. For AGW observations on the Sun, the phase shifts are often considered between fluctuations of the intensity of individual spectral lines and the vertical velocity $V_{z} $ of the particles (Deubner et al., 1990; Ichimoto et al., 1990). In the adiabatic approximation, the theory predicts $T'{\rm /}T_{0} =0$ for the f-mode, which should manifest itself in observations as a distinguishing feature of this mode. For the same reason, the Lamb mode should not be visible in the power spectrum of the vertical velocity, since that $V_{z} =0$ for it.

The study of the phase shifts can be useful for experimental diagnostics of different types of evanescent modes. Of particular interest is the possibility of observing the $\gamma $-mode, which recently obtained theoretically. As an experimental indicator of the $\gamma $-mode, the condition $\rho '{\rm /}\rho _{0} =0$ on its dispersion curve can be considered. As a result, there should be a discontinuity in the phase shift $\left(\rho '/\rho _{0} \right)-V_{z} $ when crossing dispersion curves $\omega ^{2} =k_{x} g\left(\gamma -1\right)$ and $\omega ^{2} =k_{x}^{2} c_{s}^{2} $, as follows from the polarization relation (\ref{8}).

It follows from the polarization relation (\ref{7}) that the phase shift $\left(T'/T_{0} \right)-V_{z} $ undergoes the  jump by $\pi$ if we cross both the dispersion curves for the f-mode $\omega ^{2} =k_{x} g$ and the Lamb wave $\omega ^{2} =k_{x}^{2} c_{s}^{2} $. When interpreting observations on the Sun, it is usually assumed that intensity fluctuations of Fraunhofer lines reflect the temperature fluctuations. In this case, it should be observed the jumps in the phase between the intensity and the velocity when crossing the dispersion curves of the f-mode and of the Lamb wave in the evanescence spectrum of the Sun. However, some inconsistencies of the observations with the theory were found, especially in the low-frequency region of the evanescent spectrum (Deubner et al., 1990). These discrepancies cannot be explained by taking into account the non-adiabaticity of the solar atmosphere (Deubner et al., 1990; Ichimoto et al., 1990). Perhaps, in this connection, it is necessary to take into account the dependence of the spectral line not only on the temperature, but also on the concentration of emitting particles.

The phase shift  $\left(T'/T_{0} \right)-V_{z} $ undergoes the jump at crossing the dispersion curve of the f-mode $\omega ^{2} =k_{x} g$ and the phase shift $\left(\rho '/\rho _{0} \right)-V_{z} $ changes by $\pi $ at the intersection of the curve of the $\gamma $-mode $\omega ^{2} =k_{x} g\left(\gamma -1\right)$. Therefore, both of these features can be reflected in the observed phase shift, which complicate the analysis of the diagnostic diagram.

Let us pay attention to one possible difficulty in diagnosing the $\gamma $-mode on the Sun. The observations of small wavelength disturbances on the Sun are limited by the size of photospheric granules ($\sim$1500 km), which cover the solar surface. For the f-mode, the vertical dependence of the amplitude is $a=k_{x} $, and $a=\left(1/H\right)-k_{x} $ for the $\gamma $-mode. Therefore, vertically decreasing solutions for f- and $\gamma $-modes, where the condition $a<1/2H$ is satisfied, are located in different regions of spectral plane $\left(\omega ,k_{x} \right)$. For the f-mode, the decreasing solutions are located in the lower left region of the spectral plane, where $k_{x} <1/2H$. For the $\gamma $-mode, the decreasing solutions are located in the upper right region of the diagnostic diagram, here $k_{x} >1/2H$. Consequently, the $\gamma $-mode is in the range of sufficiently small wavelengths, which complicates the possibility of its observation on the Sun due to the photospheric granulation.

\section{CONCLUSIONS}

It is shown that there are four specific evanescent wave modes in the isothermal atmosphere. These modes form the family of solutions obtaining under the condition that one of the perturbed quantities $V_{x} $, $V_{z} $, $\rho '/\rho _{0} $ or $T'{\rm /}T_{0} $ is equal to zero. In this case, from the equation system (\ref{1}) the following family of specific wave modes is obtained: 1) $\omega ^{2} =k_{x} g$, for ($V_{x} ,V_{z} ,\rho '/\rho _{0} ,0$); 2) $\omega ^{2} =k_{x} g\left(\gamma -1\right)$, for ($V_{x} ,V_{z} ,0,T'{\rm /}T_{0} $); 3) $\omega ^{2} =k_{x}^{2} c_{s}^{2} $, for ($V_{x} ,0,\rho '/\rho _{0} ,T'{\rm /}T_{0} $); 4) $\omega ^{2} =N^{2} $, for ($0,V_{z} ,\rho '/\rho _{0} ,T'{\rm /}T_{0} $). If one of the perturbed quantities vanishes, the free wave propagation becomes impossible and such the modes are horizontally propagating (evanescent).

The vanishing one of the perturbed variables is characteristic of the hydrodynamic equations written with respect to $V_{x} $, $V_{z} $, $\rho '/\rho _{0} $, $T'{\rm /}T_{0} $  (see equation (1)). If we use, for example, the equation for the disturbed pressure $p'/p_{0} $, instead of the temperature $T'{\rm /}T_{0} $, then getting the family of specific evanescent modes is more difficult.

On four dispersion curves $\omega ^{2} =k_{x} g$, $\omega ^{2} =k_{x} g\left(\gamma -1\right)$, $\omega ^{2} =k_{x}^{2} c_{s}^{2} $ and $\omega ^{2} =N^{2} $, the phase  shifts between the individual perturbed quantities $V_{x} $, $V_{z} $, $\rho '/\rho _{0} $, $T'{\rm /}T_{0} $ change abruptly by $\pi $. These phase jumps depend on the choice of a pair of the perturbed quantities. The phase shift $\left(T'/T_{0} \right)-V_{z} $ undergoes the jump when crossing the curves of the f-mode and the Lamb. For the pair of quantities $\left(\rho '/\rho _{0} \right)-V_{z} $, we obtain the phase shift after crossing both the dispersion curves of the $\gamma $-mode $\omega ^{2} =k_{x} g\left(\gamma -1\right)$ and the Lamb wave $\omega ^{2} =k_{x}^{2} c_{s}^{2} $. For the pair $\left(p'/p_{0} \right)-V_{z} $, the phase shift changes abruptly on the BV and the Lamb lines. For perturbed pair $\left(T'/T_{0} \right)-\left(\rho '/\rho _{0} \right)$ of the thermodynamic quantities, the phase shift changes if we cross the curves of the f-mode and $\gamma $-mode.

For different combinations of pairs of perturbed quantities, the phase shift between them changes abruptly by $\pi$ on those dispersion curves where one of these quantities is zero. Knowledge of these features can be useful for interpreting different types of evanescent wave modes in the observations.

The work was carried out with the support of the National Research Foundation of Ukraine, project 2020.02/0015 and partially with the support of the Target Integrated Program of the National Academy of Sciences of Ukraine for the Scientific Space Research for 2018--2022 and The Royal Society International Exchanges Scheme 2021 ''Predicting natural hazards by driven ionospheric perturbations'' (IES/R1/211177).

%\end{document}

\end{document}